\documentclass[pdftex,twocolumn,epjc3]{svjour3} 

\usepackage[T1]{fontenc}
\usepackage[utf8]{inputenc}
\usepackage[numbers,sort&compress]{natbib}
\usepackage[colorlinks,citecolor=blue,urlcolor=blue,linkcolor=blue]{hyperref}
\usepackage{hyperref}
\usepackage{xcolor}
\usepackage{amsmath}
\usepackage{latexsym}
\usepackage{amssymb}
\usepackage{graphicx}
\usepackage{mathptmx}
\usepackage{array}
\usepackage{mathtools}

\smartqed 

\journalname{Eur. Phys. J. C}

\newcommand{\mnras}{MNRAS}

\newcommand{\mtiny}[1]{{\mbox{\tiny \rm #1}}}

\newcommand{\pertorder}[2]{\prescript{\, \mtiny{(#1)}}{}{#2}}

\usepackage{xparse}
\ExplSyntaxOn
\DeclareDocumentCommand{\eqref}{m}{\quinn_mref:n {#1}}
\seq_new:N \l_quinn_mref_seq
\cs_new:Npn \quinn_mref:n #1
 {
  \seq_set_split:Nnn \l_quinn_mref_seq { , } { #1 }
  \seq_pop_right:NN \l_quinn_mref_seq \l_tmpa_tl
  ( 
  \seq_map_inline:Nn \l_quinn_mref_seq
    { \ref{##1},\nobreakspace }
  \exp_args:NV \ref \l_tmpa_tl 
  ) 
 }
\ExplSyntaxOff

\begin{document}

\title{Cosmological framework for renormalization group extended gravity at the action level}

\author{Nicolas R. Bertini\thanksref{e1,addr1}
        \and
        Wiliam S. Hip\'olito-Ricaldi\thanksref{e2,addr2}
        \and
        Felipe de Melo-Santos\thanksref{e3, addr1}
        \and \\
        Davi C. Rodrigues\thanksref{e4, addr1}
}

\thankstext{e1}{e-mail: nicolas.bertini@cosmo-ufes.org}
\thankstext{e2}{e-mail: wiliam.ricaldi@ufes.br}
\thankstext{e3}{e-mail: felipe.santos@cosmo-ufes.org}
\thankstext{e4}{e-mail: davi.rodrigues@cosmo-ufes.org}

\institute{N\'ucleo de Astrof\'isica e Cosmologia, PPGCosmo \& Departamento de F\'isica, Universidade Federal do Esp\'irito Santo. Vit\'oria, ES, Brazil.  \label{addr1}
           \and
            Departamento de Ci\^encias Naturais \& PPGCosmo, Universidade Federal do Esp\'irito Santo. S\~ao Mateus, ES, Brazil. \label{addr2}
}

\date{{}}

\maketitle

\begin{abstract}
General relativity (GR) extensions based on renormalization group (RG) flows may lead to scale-dependent couplings with nontrivial effects at large distance scales.  Here we develop further the approach in which RG effects at large distance scales are fully encoded in an effective action and we apply it to cosmology.  In order to evaluate the cosmological consequences, our main assumption is the use of a  RG scale such that the (infrared) RG effects only appear at perturbative order (not at the background level). The emphasis here is on analytical results and qualitative understanding of the implied cosmology. We employ commonly used parametrizations for describing modified gravity in cosmology (as the slip parameter). From them, we  describe the dynamics of the first order perturbations and estimate bounds on the single dimensionless parameter ($\nu$) introduced by this framework.  Possible impacts on dark matter and dark energy are discussed. It is also shown here that the $\nu$ parameter effects to $f\sigma_8$ are stronger at low redshifts ($z<1.5$), while different values for $\nu$ do not appreciably change $f\sigma_8$ at higher redshifts, thus opening a window to alleviate an issue that is currently faced by $\Lambda$CDM. 
\end{abstract}

\section{Introduction} \label{sec:intro}

Renormalization group (RG) effects to gravity at large distances (astrophysical or cosmological) are not a novelty and are being considered from different approaches (e.g., \cite{Goldman:1992qs, Bonanno:2001hi, Reuter:2003ca, Shapiro:2004ch, Borges:2007bh, Reuter:2007de, Shapiro:2008sf, Shapiro:2009dh, Nagy:2012rn,  Wetterich:2017ixo,  Eichhorn:2018yfc,    Sola:2013fka, Rodrigues:2009vf, Rodrigues:2015hba, Sola:2016jky,  Gomez-Valent:2018nib, Canales:2018tbn}). They include approaches within quantum gravity  (like asymptotically safe gravity), quantum field theory in curved spacetime (QFTCS), and phenomenological ones that emphasize the constraints from the observational data and classical symmetries. Being more specific, in the context of \linebreak QFTCS, the  Einstein-Hilbert action needs to be supplemented by higher derivative terms, in order for properly quantizing the matter sector. These higher derivative terms are dynamically relevant at small distance scales, but their importance decreases as one moves towards larges scales. Their couplings   can be shown to have trivial RG flows in the infrared, (i.e., they become true constants) \cite{Gorbar:2002pw, Gorbar:2003yt}. This behaviour is similar to the quantum electrodynamics (QED) case, where in the infrared limit the coupling can be shown to become a constant (see e.g., \cite{Goncalves:2009sk, KLOE-2:2016mgi}). However, the two other couplings of gravity, $G$ and $\Lambda$, do not need to have the same behaviour and may run in the far infrared (in this context, see e.g., \cite{Shapiro:2009dh}). Although they may run within different contexts, it is not settled how they run. The two pertinent unknowns are the $\beta$-functions, which set how the couplings depend on a RG scale $\mu$, and the relation of the latter to other physical quantities, the scale setting (e.g, \cite{Reuter:2003ca,Shapiro:2004ch, Babic:2004ev, Koch:2010nn, Koch:2014joa}).

Considering the possible running of $G$ and $\Lambda$ at astrophysical or cosmological distances, here we use two main hypothesis: $i$) at large distances, there must exist an effective description that is fully in the classical framework. In particular, there must be a complete classical action capable of effectively describing the complete large scale RG effects.\footnote{It is an extension of the improved action approach described in ref. \cite{Reuter:2003ca}. The relevant information is put in a classical action, including the meaning of the RG scale. See also ref.~\cite{Koch:2014joa}. Another well studied possibility is  implementing the RG effects at the level of the GR field equations (see e.g., \cite{Bertolami:1993mh, Reuter:2003ca, Grande:2010vg, Sola:2013fka}).   For the latter case, a complete  action is not considered, and it may even fail to exist. We add that ref.~\cite{Sola:2007sv} starts from an action and finds field equations similar to the case where the run of $G$ and $\Lambda$ are implemented at the field equations level. To this end, an external field that implements a conformal symmetry is used.} $ii$) We consider a RG scale that is essentially a measurement of the spacetime perturbations. This scale was presented in covariant form in ref.~\cite{Rodrigues:2015hba}, which can be seen as a  covariant extension of a Newtonian potential-based scale \cite{Rodrigues:2009vf, Domazet:2010bk}, which in turn extends other cases towards continuous matter distributions (e.g., \cite{Reuter:2003ca, Shapiro:2004ch}).   It is the first time that this covariant scale is applied to cosmology.  
 
For cosmology, we distinguish two classes of scales: those based on the cosmological time $t$ and those based on the perturbations wavenumber $k$. The former class  in particular includes scales that are functions of the Hubble parameter \cite{Shapiro:2000dz} (see also Refs.~\cite{Shapiro:2004ch,  Borges:2007bh, Rezaei:2019xwo}).  The latter class is the class that receives especial attention here. We remark that, for dealing with phenomena close to the singularity, it seems natural to consider the effects of the  scales based on time (and hence that change the cosmological background). However, especially for late-time cosmology, scales based on $k$ may have a relevant role. For selecting it, we also considered that: $a$) this scale is explicitly spacetime covariant (it is a scalar), which is a welcome property for inserting it in the effective action and for dealing with cosmological perturbations;  $b$) it leads to a dynamical picture that is different from GR and well-known modified gravity theories like $f(R)$ (e.g., refs.~\cite{Koch:2010nn, Hindmarsh:2012rc} show that setting the Ricci scalar as the RG scale can lead to $f(R)$); $c$) phenomenologically, considering the overall success of the standard cosmological picture, it is reasonable that possible departures are at the perturbative level. 

With this setting described above, we show that a consistent dynamical picture within cosmology requires the use of a second RG scale (multiscale RG methods are not a novelty \cite{Einhorn:1983fc, Ford:1996yc, Steele:2014dsa}). This second scale is, however, fixed from dynamical consistency. 

In what follows, we start by reviewing the action presented in ref.~\cite{Rodrigues:2015hba} and extending it towards an arbitrary number of RG scales. In section~\ref{sec:RGGRcosmology}, the consequences for gravity at cosmological scales are presented, showing that the second scale is a function of the energy-momentum tensor trace. Section~\ref{sec:CosmologicalConsequences} considers cosmological constraints and discuss the physical consequences. In section~\ref{sec:conclusions}, we present our conclusions and perspectives.

\section{Infrared renormalization group effects in gravity at the action level} \label{sec:sectionII}

Here we briefly review the approach developed in ref.~\cite{Rodrigues:2015hba}, which is especially based on 
refs.~\cite{Reuter:2003ca, Reuter:2004nx, Shapiro:2004ch, Rodrigues:2009vf}. It presents an approach in which the relevant information for implementing infrared RG effects for gravity are fully encoded in the action, instead of appending them  at the field equations level (see also \cite{Koch:2014joa} for a related approach). To put all the relevant information in the action is important for  understanding the system dynamics and symmetries (which are, independently on the underlying microphysics, seen as effectively classical at large distance scales, as explained in the introduction). Appending information at the level of the field equations is not in general equivalent to insert the information in the action and proceeding with the full variation (e.g., constrained systems). In a closed (effectively) classical environment, the existence of a complete action is expected. Also, the usefulness of an incomplete action is limited, in particular the dynamical consequences of the action diffeomorphism invariance become obscure.

In refs.~\cite{Reuter:2003ca, Reuter:2004nx, Reuter:2004nv, Shapiro:2004ch, Reuter:2007de, Rodrigues:2009vf}, it is argued in favour of the following action capable of enclosing the large scale Renormalization Group effects for gravity,
\begin{equation}
    S[g] = \frac{1}{16 \pi} \int \frac{R - 2 \Lambda}{G} \sqrt{-g}   d^4x. \label{RGexter}
\end{equation}
In the above, $G$ and $\Lambda$ are not constants, they are external scalar fields (that is, no variation with respect to either $G$ or $\Lambda$ should be considered in this action), whose running is determined from $\beta$-functions. Clearly, although this simple action has some interesting properties (e.g., \cite{Reuter:2003ca, Shapiro:2004ch, Rodrigues:2009vf}), not all relevant physical information is included in it. The dependences of $G$ and $\Lambda$ on the RG scale are not explicit, also the  physical meaning of the RG scale (the scale setting) is not in this action, these informations are appended at the level of the field equations. 

To achieve scale setting at the action level, and without recourse to external scalar fields, we use \cite{Rodrigues:2015hba},
\begin{equation}
    S  \!\! =  \! \int \left[\frac{R - 2 \Lambda(\mu)}{16 \pi G(\mu)}  + \lambda \left[ \mu - f(g,\Psi)\right]\right] \sqrt{-g} \,  d^4x   + S_\mtiny{matter}. \label{RGGRaction1}
\end{equation}
In the above, $S = S[g,\mu,\lambda,\Psi]$, $S_\mtiny{matter}=S_\mtiny{matter}[g,\Psi]$, $\Psi$ represents any additional fields, $G$ and $\Lambda$ are not external fields, both depend on the RG scale $\mu$, and the latter is seen as a fundamental field (i.e., the variation with respect to  $\mu$ is considered to find the field equations). It should be noted that $\mu$ only enters in the action \eqref{RGGRaction1} as an auxiliary field: it can be completely removed by solving  $\mu - f(g,\Psi) = 0$, as detailed in Appendix A of ref.~\cite{Rodrigues:2015hba}. We remark that this is also in agreement with RG framework expectations, in the sense that the RG scale must not be a new independent field with its own dynamics. See also ref.~\cite{Koch:2014joa} for similar arguments.

Fixing the dependence of $\Lambda$ and $G$ on $\mu$ corresponds to fixing their  $\beta$-functions. In general the action above imposes a relation between these two $\beta$-functions \cite{Rodrigues:2015hba} (see also \cite{Reuter:2003ca, Shapiro:2004ch}), this relation is indirectly related to diffeomorphism invariance and energy-momentum  conservation \cite{Rodrigues:2015hba}. If for one of them the $\beta$-function is settled considering natural arguments from the RG group, this is sufficient for fixing the other, which is found from the field equations (i.e., from the requirement that there is a consistent classical picture).

In general, RG effects need not to depend on a single scale and multiscale RG methods can be found \cite{Einhorn:1983fc, Ford:1996yc, Steele:2014dsa}. As shown in the next section, the application of the action \eqref{RGGRaction1} to cosmology may demand more than one scale, they are labeled $\mu_p$. In this case, that action can be straightforwardly extended as
\begin{align}
   & S[g,\mu,\lambda,\Psi] =  S_\mtiny{matter}[g,\Psi] + \label{RGGRaction2} \\
   & + \! \frac1{16 \pi } \int \left[\frac{R - 2 \Lambda(\mu)}{G(\mu)}  \!+\! \sum_p \lambda_p \left[ \mu_p - f_p(g,\Psi)\right]\right]\! \sqrt{-g} \,  d^4x \, .     \nonumber
\end{align}
When writing the dependences of functions or functionals, we omit the indices, thus in the above $\Lambda(\mu) = \Lambda(\mu_1, \mu_2,...)$.

From the above action, the field equations are
\begin{align}
	&{\cal G}_{\alpha \beta} + \Lambda g_{\alpha \beta}+ f_{\alpha \beta}= 8 \pi G T_{\alpha \beta} \, ,  \label{RGGRfieldEq1}\\[.1in]
	&\frac 1{16 \pi} \sum_p \int \lambda'_p \frac{\delta f'_p}{\delta \Psi}  \sqrt{-g'} \, d^4x' = \frac{\delta S_\mtiny{matter}}{\delta \Psi} \, , \label{FieldEquationsPsi}\\[.1in]
	&\mu_p - f_p = 0 \, , \\[.1in]
	&2  \frac{\partial}{\partial {\mu_p}} \frac{\Lambda}{G} -   R \frac{\partial}{\partial {\mu_p}} G^{-1}  =  \lambda_p \, , \label{consistence}
\end{align}
where a prime indicates dependence on $x'$, instead of $x$, and
\begin{eqnarray}
	{\cal G}_{\alpha \beta} &\equiv & G_{\alpha \beta} +   G \Box G^{-1} g_{\alpha \beta} - G \nabla_\alpha \nabla_\beta G^{-1} \, \label{Gcaldef},\\
	f_{\alpha \beta} &\equiv &  - \frac {G} {\sqrt{-g}} \sum_p \int \lambda'_p \frac{\delta f'_p}{\delta g^{\alpha \beta}} \sqrt{-g'} \,  d^4 x' \, \label{fabdef},\\
	T_{\alpha \beta} &\equiv & -  \frac 2 {\sqrt{-g}} \frac{\delta S_\mtiny{matter}}{\delta g^{\alpha \beta}} \, .\label{Tdef}
\end{eqnarray}
To express the field equations above, we used functional derivatives. In particular, for some field $\phi$ and function $f = f(\phi, \partial \phi)$, 
\begin{align}
&  \frac{\delta \phi'}{\delta \phi} \equiv  \frac{\delta \phi(x')}{\delta \phi(x)} = \delta^{(4)}(x-x') \, ,\\
&	\frac{\delta f'}{\delta \phi} = \frac{\partial f}{\partial \phi}(x) \; \delta^{(4)}(x-x') + \frac{\partial f}{\partial (\partial_\alpha \phi)}(x) \; \partial'_\alpha \delta^{(4)}(x-x') \, .
\end{align}

From the diffeomorphism invariance of  $S_\mtiny{matter}$ (e.g., \cite{Wald:1984rg}), 
\begin{align}
	0 &= \delta_\xi S_\mtiny{matter}[g, \Psi]  \nonumber \\
	&=  \int \left( -\frac 12 T_{\alpha \beta} \sqrt{-g} \nabla^\alpha \xi^\beta + \frac{\delta S_\mtiny{matter}}{\delta \Psi} \delta_\xi \Psi \right)d^4 x \nonumber \\[.1in]
	&=  \int \left( \frac 12 \nabla^\alpha T_{\alpha \beta} \sqrt{-g} \,  \xi^\beta  + \right.  \nonumber \\ 
	& \;\;\; \left. + \frac 1 {16 \pi} \sum_p \int \lambda'_p \frac{\delta f'_p}{\delta \Psi}  \sqrt{-g'} \, d^4x' \, \, \delta_\xi \Psi \right)d^4 x \, ,    \label{diff}
\end{align}
where $\delta_\xi$ represents an infinitesimal change of coordinates, given by a Lie derivative along the vector $\xi^\alpha$. Hence, a violation of energy-momentum tensor conservation requires that some $\lambda_p$'s are not zero.

In ref.~\cite{Rodrigues:2015hba}, simpler systems were considered than cosmology with perturbations. There, a single scale was sufficient  and the selected RG scale was such that the single Lagrangian multiplier ($\lambda$) was zero at the field equations level, thus implying $\nabla_\alpha T^{\alpha \beta} =0$. As it will be shown, for cosmology a modest violation of energy-momentum conservation will appear, since we will find $\lambda_2 \not = 0$.

\section{Gravity and matter at cosmological scales} \label{sec:RGGRcosmology}
\subsection{Spacetime metric} \label{sec:spacetimemetric}
Similarly to standard cosmology,  it is assumed that spacetime can be  foliated and that the universe at large scales can be described by a spatially homogeneous and isotropic metric, added by non-homogenous perturbations. Only  scalar perturbations are considered here, since they are the most relevant for the large scale structure. For clarity,  considering that the main purpose of the present work is to establish the cosmological framework, the spatial slices are taken to be flat. Hence, the line element in the Newtonian gauge can be written as
\begin{equation}
	ds^2 \!=\! -a^2(\eta) (1 + 2 \psi) d\eta^2	\!+\! a^2(\eta) (1 - 2 \phi) \delta_{ij} dx^i dx^j\, , \label{lineelement}
\end{equation}
where $\eta$ is the conformal time, $a$ is the scale factor, $\psi$ and $\phi$ are the first order metric perturbations.

The cosmological background is written as
\begin{equation}
	\pertorder{0}{g}_{\alpha \beta} = a^2(\eta) \, \eta_{\alpha\beta} \,,
\end{equation}
where $\eta_{\alpha \beta}$ is the Minkowski metric.

\subsection{The main RG scale} \label{sec:RGbackground}

We consider, as commented in the introduction, a RG scale that can be written as a scalar, such that it is possible to explicitly insert it in the action, and we consider the case in which the scale is directly connected to the cosmological perturbations (i.e., to the wavenumber scale $k$). This choice  is motivated from three considerations: $i$) there is already a candidate for such scale that satisfies these conditions, which is the scale proposed in ref.~\cite{Rodrigues:2015hba} and detailed further below; $ii$) it is a possibility less explored in cosmology, which we think deserves further attention;  $iii$) $\Lambda$CDM is facing some difficulties but it has already achieved relevant success on a large class of cosmological and astrophysical phenomena. Therefore, cosmological models whose background field equations are the same of $\Lambda$CDM, but with differences at the perturbative level, sounds worth to investigate. 
 
Although the proposal from ref.~\cite{Rodrigues:2015hba} can be seen as an extension of various other proposals, in particular of the proposal for galaxies in refs.~\cite{Shapiro:2004ch, Reuter:2004nx, Rodrigues:2009vf}, the cosmological model of ref.~\cite{Shapiro:2004ch}, and some other works with similar field equations (e.g., \cite{Sola:2007sv, Grande:2010vg}), is different from the one proposed here. Bounds from other cosmological models, independently on the $\beta$-function adopted for $G$, do not apply to the present case. As an example, big bang nucleosynthesis (BBN) bounds were evaluated for a cosmological model with RG effects in ref.~\cite{Grande:2010vg}, but these bounds do not apply to the present case since here the background equations are not sensitive to RG effects.
 
The matter content is taken to be a perfect fluid whose 4-velocity is denoted by $U^\alpha$.

The main RG scale ($\mu_1$) is set to be equal to a function of the scalar $W$. This scalar was introduced in ref.~\cite{Rodrigues:2015hba} and is a function of  $U^\alpha$, the metric $g_{\alpha \beta}$ and certain tensor denoted by $\gamma_{\alpha \beta}$, as follows:
\begin{align} 
		& \mu_1 = f_1(W) \, ,  \label{mu1scale} \\
		& W \equiv U^\alpha U^\beta (g_{\alpha \beta} - \gamma_{\alpha \beta}) \, .\label{Wdef}
\end{align}
One can note that $W$ is a scalar measurement of the difference $g_{\alpha \beta} - \gamma_{\alpha \beta}$ and $\gamma_{\alpha \beta}$ provides the reference geometry to that difference. To fulfill this interpretation as a reference, $\gamma_{\alpha \beta}$ should not have a kinetic term and it should only appear inside $W$, which is assumed henceforth. These steps are all in accordance with ref.~\cite{Rodrigues:2015hba}.

The equations above particularize the general dependence of $f_1$ from $f_1=f_1(g, \Psi)$ into $f_1=f_1(W(g, U,\gamma))$.  Since the setting of $\mu_1$ is done at the action level, the field equations depend in general on the variation of $f_1$ with respect to the $\Psi$ fields \eqref{RGGRfieldEq1, FieldEquationsPsi, fabdef}. This feature is absent from other RG implementations in which the scale setting is done at the field equations level, but we understand it as  a necessity if one looks for an action with all the dynamical information in it (including the scale setting).

The tensor $\gamma_{\alpha \beta}$ enters the action \eqref{RGGRaction2} as a fundamental field, being one of the fields that compose the $\Psi$ set of fields ($U^\alpha$ is another field that is also part of $\Psi$). Before continuing, we display here the current structure of the action. Let $\gamma_{\alpha \beta}$ be a field that only appears inside $f_1$; and let $\hat \Psi$ be a set of fields that include all the $\Psi$ fields except for $\gamma_{\alpha \beta}$ (i.e., $\delta \hat \Psi/ \delta \gamma_{\alpha \beta} = 0$). Therefore, the action \eqref{RGGRaction2} can be rewritten as
\begin{align}
   & S[g,\mu,\lambda, \gamma,\hat \Psi] =  S_\mtiny{matter}[g,\hat \Psi] +  \! \frac1{16 \pi } \int \left[\frac{R - 2 \Lambda(\mu)}{G(\mu)}   \right. + \nonumber \\[.2cm]
   & \left. \!+\! \lambda_1 \left[ \mu_1 - f_1(W) \right] \!+\! \sum_{p=2} \lambda_p \left[ \mu_p - f_p(g,\hat \Psi)\right]\right]\! \sqrt{-g} \,  d^4x \, .     \label{RGGRaction3}
\end{align}
We stress that  $W$ is not a fundamental field in this action, it a function of $g_{\alpha \beta}$, $\gamma_{\alpha \beta}$ and $U^\alpha$ (i.e., $W = W(g, U, \gamma)$), and $U^\alpha$ is one of the fields that is part of the set $\hat \Psi$.

A relevant consequence of using $\gamma_{\alpha \beta}$ as a fundamental action field, instead of an external one, is that, the variation with respect to $\gamma_{\alpha \beta}$ implies that $\lambda_1 =0$ (at the level of the field equations). This is shown explicitly in the next subsection. This implies that setting either the scale at the level of the action or at the level of the field equations leads to the same field equations. The latter statement is valid for the scale $\mu_1$. This is why the field equations of ref.~\cite{Rodrigues:2015hba}, which considers a scale setting at the level of the action, are compatible with ref.~\cite{Rodrigues:2009vf} equations. 

In the context of local structures (e.g., solar system, a galaxy...), a natural choice for $\gamma_{\alpha \beta}$ would be the Minkowski metric ($\eta_{\alpha \beta}$), such that, far from the system $g_{\alpha \beta}$ asymptotically becomes Minkowski and $W$ asymptotically becomes zero.  Hence, in this context and in a comoving frame with the system, $W$ can be written as the metric time-time component perturbation, $W \stackrel{*}= U^0 U^0( g_{00} - \eta_{00}) \approx  g_{00} - \eta_{00}$. Therefore, $W$ is the Newtonian potential computed in a comoving frame (apart from a factor 2 and higher order corrections). Such Newtonian potencial choice was used in the context of galaxies and the solar system \cite{Rodrigues:2009vf, Farina:2011me, Domazet:2010bk, Rodrigues:2012qm, Toniato:2017wmk} (besides a star-like case \cite{Rodrigues:2015hba}). It also constitutes an extension of some others scale settings considered in the context of a point particle (e.g., \cite{Reuter:2003ca, Shapiro:2004ch, Reuter:2004nx}). 

In a cosmological context, and in accordance with the motivation of using a RG scale mainly based on the wavenumber $k$ of the perturbations, instead of the time $t$ scale (section~\ref{sec:intro}), we consider solutions in which $\gamma_{\alpha \beta}$ is the cosmological background, i.e.,
\begin{equation} \label{eqbackgrounds}
	\gamma_{\alpha \beta} = \pertorder{0}{g}_{\alpha \beta} \, .
\end{equation} 
Therefore, using the line element \eqref{lineelement},\footnote{To stress that the computation is done in a particular coordinate system, the comoving frame (defined by $U^i =0$), we use the symbol ``$\stackrel*=$''.}
\begin{equation} \label{Wcosmo}
	W \stackrel*= \frac 1{a^2} (g_{0 0} - \pertorder{0}{g}_{0 0 }) = - 2 \psi  \, .
\end{equation}
This is the same scale used in the other works cited above, being in essence the Newtonian potential. 

In the approach that we are dealing here, it will not be necessary specify the function $f_1$ further, the important step is to state its dependence, as  in eq.~\eqref{mu1scale}. With this setting, RG effects will be sensitive and change the cosmological perturbations, but the background field equations will not depend on them.

\subsection{Background equations}

If  $W=0$ in a given spacetime region, then  $G$ and $\Lambda$ should have no RG corrections, that is they should be constants in that region. Hence, let 
\begin{eqnarray}
	G|_{W=0} = G_0 \, \mbox{ and } \, \Lambda|_{W=0} = \Lambda_0 \, . 	 \label{LambdaBC} 
\end{eqnarray}
The  above will be used as boundary conditions that lead to unique solutions in vacuum.

Since the  action dependence on $\gamma_{\alpha \beta}$ is only through $f_1$ \eqref{mu1scale}, this implies, from the field equations, that $\lambda_1 = 0$ at the level of the field equations \cite{Rodrigues:2015hba}. Indeed, using $\gamma_{\alpha \beta}$ in place of $\Psi$ in eq.~\eqref{FieldEquationsPsi}, the right hand side is zero, therefore,
\begin{align}
	  0= \int \lambda_1' \frac{\delta f_1'}{\delta \gamma_{\alpha \beta}} \sqrt{- g'} d^4x'  =   \lambda_1 \frac{\partial f_1}{\partial W} U^\alpha U^\beta \sqrt{- g}   \, .
\end{align}
Since we assume $\frac{\partial f_1}{\partial W}\not=0$ and since the other quantities cannot be zero,  the solution is 
\begin{equation}
	 \lambda_1  =	0	\, . \label{lambda1sol}
\end{equation}

Using $h_{\alpha \beta}$ to denote the metric perturbations,
\begin{equation}
	h_{\alpha \beta} \equiv  g_{\alpha \beta} - \pertorder{0}{g}_{\alpha \beta} \, ,
\end{equation}
the background equations can be found by neglecting all the contributions of first or higher orders on $h_{\alpha \beta}$. In this regime, eq.~\eqref{consistence}  with eqs.~(\ref{LambdaBC}, \ref{lambda1sol}) imply that all the Lagrange multipliers are zero at background level ($\lambda_p \approx 0$). Consequently, the matter field equations, eq.~\eqref{FieldEquationsPsi}, becomes simply  $\delta S_\mtiny{matter} / \delta \Psi \approx 0$. Moreover, since $f_{\alpha \beta} \approx 0$ and ${\cal G}_{\alpha \beta} \approx G_{\alpha \beta}$, eq.~\eqref{RGGRfieldEq1} becomes Einstein field equations. This completes the verification that at background level there are no RG corrections in this framework (in the sense that the form of the equations is the same of GR at background level). We stress that to achieve this result  we used the scalar $W$ as the RG scale. It is also relevant to stress  that this result has no dependence on how $G$ and $\Lambda$ depend on the scales $\mu_p$ (the $\beta$-functions), neither on the precise form of the functions $f_p$ (scale settings), apart from $\mu_1 = f_1(W)$. A second RG scale will also be necessary, but it will not (and cannot) change this result.

Below, we write down the background equations, which are Friedmann equations with background energy density $\pertorder{0}{\epsilon}$ and pressure $\pertorder{0}{p}$,
\begin{align}
	&3 {\cal H}^2 - \Lambda_0 a^2 = 8 \pi  G_0 \, a^2 \pertorder{0}{\epsilon} \, , \label{friedmann1}\\
	& 2 {\cal H}' +  {\cal H}^2 -  \Lambda_0 a^2 = -8 \pi  G_0 \, a^2  \pertorder{0}{p} \, , \label{friedmann2}
\end{align}
where ${\cal H} \equiv a'(\eta)/a(\eta)$ is the Hubble parameter in conformal time. A prime here  denotes derivative with respect to conformal time.

\subsection{The relation between $G$ and $\Lambda$ in vacuum} \label{sec:GLvacuum}

The framework, as presented up to this point, is sufficient for deriving this relation. The generalization towards many RG scales, as done in this work, does not change the relation derived in \cite{Rodrigues:2015hba} (which assumes a single RG scale). The explicit form of the $G$ an $\Lambda$ relation, to be shown below, opens a possible interpretation as a consequence of the existence of a IR fixed point in the RG flow, as commented below. 

Since at background level GR is valid, then, in vacuum ($T_{\alpha \beta} = 0$),
\begin{equation}
	\pertorder{0}{R}  = 4 \Lambda_0 \, .	
\end{equation}
For any quantity $X$, $\pertorder{0}{X}$ means the background value of $X$.

Therefore, from eq.~\eqref{consistence} and  up to the first order on $h_{\alpha \beta}$, 
\begin{equation}
		\frac{\partial}{\partial \mu_1} \Lambda =  \Lambda_0  G_0 \frac{\partial}{\partial \mu_1} G^{-1} \, . \label{consistenceFORmu1} 
\end{equation}
The general solution of the above equation, with \eqref{LambdaBC} as the boundary conditions, reads
\begin{equation}
	\Lambda = \Lambda_0 G_0 G^{-1} \, . \label{LGrunningvac}
\end{equation}
Inserting the above solution back into eq.~\eqref{consistence}, but considering other values for $p$, one only concludes that, in vacuum, $\lambda_p = 0$. Hence, without changes to the above solution.

\subsection{The relation between $G$ and $\Lambda$ in the presence of matter} \label{sec:GLwithmatter}
Here the presence of an energy-momentum tensor $T_{\alpha \beta}$ will be considered. From eq.~(\ref{consistence}), and using the background equations, up to the first perturbative order, 
\begin{equation}
			\partial_{\mu_{p}} \Lambda =   \xi   G_0 \partial_{\mu_p} G^{-1} +  \frac 12 G_0 \lambda_p \, , \label{consistenceXi} 
\end{equation}
with
\begin{equation}
	\xi \equiv   \Lambda_0 - 4 \pi G_0 \pertorder{0}{T}. \label{xidef}
\end{equation}
As defined above, $\xi$ is a background quantity.  Since $\mu_1$ is a function of $W$ (which is a spacetime function), while $\xi$ only depends on time, they are independent. On the other hand, $\Lambda$ cannot be simply written as a function of $\mu_1$ and time $\eta$, it should only depend on $\mu_p$. Otherwise, it would not be compatible with the action \eqref{RGGRaction2}. In order to be able to express $\Lambda$ as a $\mu_p$ function, eq.~\eqref{consistenceXi} is showing that $p$ cannot assume a single value, additional RG scales beyond the first one will be necessary. For $p=1$,  recalling that $\lambda_1=0$ and recalling the boundary condition \eqref{LambdaBC}, it is possible to integrate eq.~\eqref{consistenceXi} and find
\begin{equation}
	\Lambda = \Lambda_0 +  \xi \,  \delta_G \, , \label{Lsolution}
\end{equation}
with
\begin{equation}
	\delta_G \equiv  G_0 G^{-1} -1 \, .	\label{deltaGdeff}
\end{equation}

Inserting the result above back into eq.~\eqref{consistenceXi}, with $p=2$, one finds  that  
\begin{equation}
	\xi \partial_{\mu_2} \delta_G + \delta_G \, \partial_{\mu_2} \xi  = \xi \partial_{\mu_2} \delta_G\, + \frac 12 G_0 \lambda_2 \, .
\end{equation}
Therefore,
\begin{equation}
		 \lambda_2 =  2  G_0^{-1}  \delta_G \partial_{\mu_2} \xi  \, . \label{lambda2sol}
\end{equation}
The above indicate that $\mu_2$ should be seen as a  function of the background quantity $\xi$. Thus, we let
\begin{equation}
	\mu_2 = f_2(\xi  \, ) \, . \label{mu2scale}
\end{equation} 
Equivalently, one could state that $\mu_2$ is a function of $\pertorder{0}{T}$. It will be shown shortly that the precise form of the $f_2$ function is irrelevant, the important statement is that $f_2$ is a function of $\xi$ alone.

Since all the $\mu_p$ RG scales are assumed to be independent among themselves (e.g., $\partial_{\mu_3} \mu_2 =0$),  from eqs.~(\ref{lambda2sol}, \ref{consistenceXi}) one concludes that
\begin{equation}
	\lambda_p = 0 \;\;\;\;  \forall \, p \not= 2 \, . \label{lambdapsol}
\end{equation}

Although it is not impossible to introduce new independent scales, eq.~\eqref{Lsolution} is a clear statement that two scales are sufficient in this context.

In conclusion, the solution for the relation between $\Lambda$ and $G$ is given by eq.~\eqref{Lsolution}, which shows that  in general $\Lambda$ is not a function of $G$ alone: it also depends on the matter fields through $\pertorder{0}{T}$. The set of scales $\mu_1$ and $\mu_2$, eqs.~(\ref{mu1scale}, \ref{mu2scale}) is sufficient for a consistent  derivation of the $\Lambda$ and $G$ running. Here the general solution for all the Lagrange multipliers $\lambda_p$ was also found.

\subsection{Field equations and perfect fluids} \label{sec:fieldfluid}

With the above, we have found solutions for $\Lambda$ and $\lambda_{ p}$. These can be inserted in the field equations (\ref{RGGRfieldEq1}, \ref{FieldEquationsPsi}) to yield
\begin{align}
	& {\cal G}_{\alpha \beta} + (\Lambda_0 +\xi \, \delta_G) g_{\alpha \beta} -\nonumber \\ 
	& \;\;\; - \frac 2{\sqrt{-g}} \int \delta_G'    \frac{\partial \xi'}{ \partial \mu_{2}'} \frac{ \delta f'_{2}}{ \delta g^{\alpha \beta}}	 \sqrt{-g'}\, d^4x'= 8 \pi G T_{\alpha \beta} \, ,  \label{RGGRfieldEq1b}\\[.1in]
	& \frac{\delta S_{\mtiny{matter}}}{\delta \Psi} = \frac 1{8 \pi G_0} \int \delta_G' \,    \frac{\partial \xi'}{ \partial \mu'_{2}} \frac{ \delta f'_{2}}{ \delta \Psi}   \sqrt{-g'} \, d^4x' \, . \label{FieldEquationsPsib}
\end{align}
The primes inside integrals denote spacetime dependence on $x'$, instead of $x$. Since $f_2$ is a function of $\pertorder{0}{T}$, in general it can depend on both the matter fields and the metric. However, for the perfect fluid case $\xi$ is simply a function of the energy density and the pressure, thus the term $\delta f'_{2} /  \delta g^{\alpha \beta}$ is zero. Moreover, $f_2(\xi)$ has no dependence on spacetime derivatives, hence $\delta f'_2 / \delta \Psi = \partial f_2 / \partial \Psi \, \delta(x-x')$. Therefore, for the perfect fluid case,
\begin{align}
	& {\cal G}_{\alpha \beta} + (\Lambda_0 +\xi \, \delta_G) g_{\alpha \beta} = 8 \pi G T_{\alpha \beta} \, ,  \label{RGGRfieldEq1bSimp}\\[.1in]
	&  \frac{\delta S_\mtiny{matter}}{\delta \Psi} = \frac 1{8 \pi G_0}  \delta_G \,    \frac{\partial \xi}{ \partial \Psi}   \sqrt{-g}\, . \label{FieldEquationsPsibSimp}
\end{align}
In the above, there is no dependence on the form of the $f_2$ function, we only used that $f_2$ is a (differentiable) function of $\xi$.

At background level there are no RG correction in this framework, hence, for a perfect fluid, $\pertorder{0}{T} = -\pertorder{0}{\epsilon} + 3 \pertorder{0}{p}$, where $\pertorder{0}{\epsilon}$ and $\pertorder{0}{p}$ are the energy density and the pressure at background level.  This implies that, adopting a thermodynamic description based on the mass density $n$ and the specific entropy $s$, then $\xi$ is at most a function of $n$ and $s$:
\begin{equation}
	\xi = \xi(n,s)\, . \label{xins}
\end{equation} 
In particular, for a dust-like fluid, $\xi$ only depends on $n$, while for a radiation fluid $\xi$ depends on neither of them and it is a constant.

Since $\partial \xi /  \partial \Psi$ can be different from zero,  the energy-momentum tensor derived from the matter part alone will not be conserved. This is a well-known possibility in the context of varying $G$ and $\Lambda$ (e.g., \cite{Sola:2013fka, Bonanno:2017pkg}). Nonetheless, as it will be shown, for the present case it is an especially mild type of violation.

In order to better understand the consequences of this approach in the presence of matter,  we consider a specific matter action which models an arbitrary relativistic perfect fluid \cite{1972JMP....13.1451R},
\begin{align}   
   &S_\mtiny{fluid} = \int \left[-\epsilon(n,s) + \eta_1 (1 + U^\alpha U_\alpha) + \eta_2 \nabla_\alpha (n U^\alpha) + \right. \nonumber \\ 
    &\;\;\;\; \left.+ \eta_3 U^\alpha  \nabla_\alpha X +  \eta_4 U^\alpha  \nabla_\alpha s \right]\sqrt{-g}\,d^4x \, .       \label{eq:Sfluid}
\end{align}
In the above, $S_\mtiny{fluid}  = S_\mtiny{fliud} [g, U, n, s, \eta_m,X]$, $n$ is the fluid mass density, $s$ the rest specific entropy, $\eta_m$ stands for the four Lagrange multipliers and $\epsilon(n,s)$ is the energy density. The quantity $X$ is relevant for the description of fluids with rotational flow \cite{1972JMP....13.1451R}. There are other equivalent action formulations capable of describing an arbitrary perfect fluid, but we find the action above suitable for this application since in this formulation $U^\alpha$ enters as a fundamental action field.

The energy-momentum tensor \eqref{Tdef} of this fluid is directly found as
\begin{equation}
	T_{\alpha \beta}  =  2 \eta_1 U_\alpha U_\beta +  g_{\alpha \beta} (- n U^\alpha \partial_\alpha \eta_2 - \epsilon) \, .
\end{equation}
In the above, some of the constrains inferred from the action \eqref{eq:Sfluid} variation with respect to  $\eta_m$ were used.

From eqs.~(\ref{FieldEquationsPsibSimp}, \ref{eq:Sfluid}) and  using either $n$ or $U^\alpha$ in place of $\Psi$, one finds respectively
\begin{align}
	&  \partial_n\varepsilon + U^\alpha \partial_\alpha \eta_2  = -\frac 1{8 \pi G_0} \delta_G \,  \partial_n \xi = \frac 12 \delta_G \, \partial_n \pertorder{0}{T}   \, ,\\
	& 2 \eta_1 U_\alpha - n \partial_\alpha \eta_2 = 0\, .
\end{align}

Using the constraint $U^\alpha U_\alpha = -1$, the above equations can be used to eliminate $\eta_1$ and $\eta_2$, leading to
\begin{align}
	T_{\alpha \beta} & =  (n \partial_n \epsilon - \frac 12  \delta_G	\, n\partial_n \pertorder{0}{T} ) U_\alpha U_\beta + \nonumber \\
    & \;\;\;\; +  g_{\alpha \beta} (n \partial_n \epsilon - \frac 12  \delta_G	\, n\partial_n \pertorder{0}{T}  - \epsilon)  \nonumber \\
	& =   (\epsilon_\mtiny{eff} + p_\mtiny{eff}) U_\alpha U_\beta +  g_{\alpha \beta} p_\mtiny{eff},
\end{align}
where
\begin{eqnarray}
		p_\mtiny{eff} & = & p  + \frac 12  \delta_G	\, n\partial_n (\pertorder{0}{\epsilon} - 3 \pertorder{0}{p} )\, , \label{pressurechange}\\
		\epsilon_\mtiny{eff} & = & \epsilon \, ,
\end{eqnarray}
with $p = n \partial_n \epsilon - \epsilon$ \cite{1972JMP....13.1451R}. 

The effective (``eff'') quantities are such that the position that they occupy in $T_{\alpha \beta}$ are the usual ones.  The dynamical impact of these quantities are analysed in the next subsection.

\subsection{Equations of motion in the presence of a relativistic fluid} \label{sec:eomrel}

For GR, diffeomorphism invariance implies that $\nabla^\alpha T_{\alpha \beta} =0$, and this vector equation determines the equations of motion. For the present context with $G$ and $\Lambda$ running, diffeomorphism invariance of the matter action imply the general relation presented in eq.~\eqref{diff}. For the relativistic perfect fluid case, using eqs.~(\ref{xidef}, \ref{lambda2sol}, \ref{lambdapsol}, \ref{xins}), it can be expressed as 
\begin{eqnarray}
	\nabla^\alpha T_{\alpha \beta} &=& -\frac 1{4 \pi G_0} \delta_G \left( \frac{\partial \xi}{\partial n} \partial_\beta n  + \frac{\partial \xi}{\partial s} \partial_\beta s  \right) \nonumber \\
	&=&  \delta_G \frac{\partial \pertorder{0}{T}}{\partial n} \partial_\beta n \, . \label{divTdiff}
\end{eqnarray} 
In the above, it was also used that $\delta_\xi \Psi = \xi^\alpha \partial_\alpha \Psi$ (for $\Psi$ equal to either $n$ or $s$) and that $\pertorder{0}{T}$ do not depend on $s$. As an example, for a pressureless fluid at the background level (which implies $\pertorder{0}{\epsilon} \propto \pertorder{0}{n}$), the above expression becomes, up to first order, 
\begin{equation}
	\nabla^\alpha T_{\alpha \beta} = - \, \delta_G  \, \partial_\beta  \pertorder{0}{\epsilon} \, .	\label{Qmatter}
\end{equation}

In general, and up to the first order, one can write eq.~\eqref{divTdiff} as
\begin{equation}
	\nabla_\alpha T^{\alpha \beta} =  Q^\beta \, , \label{Qvectordef}
\end{equation}
where $Q^\beta$ is a first order quantity whose single non-null component is the zeroth one. 

In a frame that is comoving with the fluid at background level, the fluid equations can be written in a form that is independent from $Q^\beta$, up to the first order,  as we show below. The previous equation, for an effective perfect fluid, can be written as
\begin{eqnarray}
	\nabla_\alpha \left ( (\epsilon_\mtiny{eff} + p_\mtiny{eff} ) U^\alpha U^\beta + g^{\alpha \beta} p_\mtiny{eff} \right ) = Q^\beta	\, . \label{divFluidEff}
\end{eqnarray}
Multiplying by $U_\beta$,
\begin{eqnarray}
	- \nabla_\alpha \left [ (\epsilon_\mtiny{eff} + p_\mtiny{eff} ) U^\alpha \right ] + \frac{D p_\mtiny{eff}}{D \tau} = U_\beta Q^\beta = U_0 Q^0 \, ,
\end{eqnarray}
where, for any quantity $X$, $D X / D\tau \equiv U^\alpha \nabla_\alpha X$. Inserting this result into eq.~\eqref{divFluidEff},
\begin{eqnarray}
	U^\beta  \frac{D p_\mtiny{eff}}{D \tau}  + (\epsilon_\mtiny{eff} + p_\mtiny{eff} ) \frac{D U^\beta}{D \tau} + \nabla^\beta p_\mtiny{eff} = Q^\beta +  U^\beta U_0 Q^0\, .
\end{eqnarray}
The above equation is the same one that can be found from GR for a fluid with energy density $\epsilon_\mtiny{eff}$ and pressure $p_\mtiny{eff}$, apart from the limit $Q^\beta \rightarrow 0$. However, the previous limit is not even necessary, since the right hand side is already zero up to the first order. This can be directly checked by considering the cases $\beta =0$ and $\beta = i$. Therefore, up to the first order,
\begin{eqnarray} \label{geod}
		 (\epsilon_\mtiny{eff} + p_\mtiny{eff} ) \frac{D U^\beta}{D \tau} + \nabla^\beta p_\mtiny{eff} + U^\beta \frac{D p_\mtiny{eff}}{D \tau} = 0 \, ,
\end{eqnarray}
just like a standard relativistic fluid. In particular, for $p_\mtiny{eff} =0$, one finds the geodesic equation ${D U^\beta}/{D\tau} =0$. It is important to stress that these are first order results which hold in any frame that is comoving with the fluid at the background level.

The results above show that dynamically, in a comoving frame, $\epsilon_\mtiny{eff}$ and $p_\mtiny{eff}$ have the same role of $\varepsilon$ and $p$ in theories with $\nabla_\alpha T^{\alpha \beta}=0$. Hence, in systems in which the fluid equation of state is fixed from the phenomenology, only $\epsilon_\mtiny{eff}$ and $p_\mtiny{eff}$ are relevant, since the difference between $p$ and $p_\mtiny{eff}$ cannot be measured independently. Whereas, in physical situations in which the equation of state is assumed to be known independently from gravitational effects, the relation between $\varepsilon$ and $p$ is known beforehand thus the pressure change \eqref{pressurechange} should be considered. Independently on the case, for a radiation fluid the effective and the fundamental pressure are always equal.

In the following, taking in consideration clarity and simplicity, we develop cosmology based on the effective pressure, not the fundamental one. That is, a dust fluid is such that $p_\mtiny{eff}=0$. It would be interesting to look as well for the case based on the fundamental pressure $p$, but we let this  case for a future work. Since only the effective quantities will be used, to simplify the notation the ``eff'' with $\epsilon$ and $p$ will no longer be used: $\varepsilon_\mtiny{eff} \rightarrow \varepsilon$ and $p_\mtiny{eff} \rightarrow p$.

\section{Cosmology and physical bounds} \label{sec:CosmologicalConsequences}

\subsection{Equations for the scalar perturbations and the slip parameter} \label{sec:eqsslip}

In order to proceed towards cosmology, it will be relevant to particularize the $\beta$-function of $G$, which states $G$ as a function of the scale $\mu_1$,  and the scale setting which express $\mu_1$ as a function of other physical quantities. Instead arguing in favour of particular realizations of each step individually, we consider an  approach that includes a relevant class of functions for exploring this framework for small $W$ values. Namely, we consider that the combination of the two previous steps leads to an analytical function about $W=0$, thus implying that
\begin{equation}
	G_0 G^{-1}(W) = 1 +   \nu W  + O(W^2)\, , \label{GWexpansion}
\end{equation} 
where $\nu$ is a dimensionless constant that can be either positive or negative and parametrizes the amplitude of the running of $G$.   We point out that the expression \eqref{GWexpansion} is quite general for small $W$ values in the sense that we only demanded $G(W)$ to be compatible with a linear expansion about $W=0$. Nonetheless, there is a relevant case that is not explicitly included in the expansion above, which will be commented latter on.

Therefore, using eqs.~\eqref{Wcosmo, deltaGdeff}, in a comoving reference frame,
\begin{align}
	& G_0 G^{-1}(W) \stackrel*= G_0 G^{-1}(\psi) = 1 - 2 \nu \psi + O(\psi^2)\, ,	\label{GPhiexp} \\
	& \delta_G \stackrel*\approx - 2 \nu \psi\, . \label{deltaGphi}
\end{align}

There are several works that consider $\beta$-functions that lead to logarithmic running for $G(\mu_1)$ (e.g., \cite{Fradkin:1981iu, Nelson:1982kt, Shapiro:2004ch, Bauer:2005rpa, Sola:2007sv}). Some of these realizations can be captured by the linear expansion above. For instance, let
\begin{equation}
	G_{\rm ln}(\mu_1)\equiv \frac{G_0}{1 + 2 \nu \ln \mu_1}\, ,\label{Gln}
\end{equation}
From this particular $G(\mu_1)$ realization, the choice $\mu_1 = W$ is not viable, since $G_{\rm ln}$ would not be finite at background level ($W=0$), thus the condition \eqref{LambdaBC} would not be possible. Equations \eqref{GWexpansion, GPhiexp} can be found from  the setting $\mu_1 = 1 + \frac 12 W \stackrel*= 1 - \psi$ and up to first order on $\psi$.  Any other linear relation between $\mu_1$ and $W$ (with non-zero constant term) is viable, and they differ by a straightforward rescaling on the $\nu$ and $G_0$ constants. Quadratic or higher order corrections on $W$ can be assumed, but they are irrelevant for the linear cosmological perturbations. Considerations about using $\mu = 1 - \psi$ together with $G_{\rm ln}$ can also be found in refs.~\cite{Rodrigues:2016tfm,  Toniato:2017wmk}. We stress that all the results here presented do not depend on eq.~\eqref{Gln}, this equation appears here just as a relevant particular case.

See also ref.~\cite{Toniato:2017wmk} for a similar expression. There are, nonetheless, interesting cases not covered by the expression above, which will be commented latter.

From eqs.~\eqref{xidef, Lsolution, deltaGphi},   $\Lambda$ is found to be
\begin{equation} 
	\Lambda = \Lambda_0 + \delta{\Lambda} = \Lambda_0 + \left( 4 \pi G_0 \pertorder{0}{T} - \Lambda_0   \right) 2  \nu \psi\,. \label{L1}
\end{equation}

For the energy-momentum tensor, we use a perfect fluid with
\begin{equation}
	T^{\alpha \beta} = \pertorder{0}{T^{\alpha \beta}} + \delta{T^{\alpha \beta}} = \varepsilon U^\alpha U^\beta + p (g^{\alpha \beta} + U^{\alpha} U^{\beta})\, .
\end{equation}
The spatial velocity perturbations are written as $\vartheta^i$, and the energy density and pressure are expanded as $\varepsilon = \pertorder{0}{\epsilon} + \delta \varepsilon$ and $p = \pertorder{0}{p} + \delta p$.

From the above expression for $G$ and $T_{\alpha \beta}$, considering the line element \eqref{lineelement} and field equations \eqref{RGGRfieldEq1bSimp}, the first order differential equations for the perturbations read 
\begin{align}
&3\mathcal{H}({\phi}'+\nu \psi') -\nabla^{2}(\phi+ \nu\psi)+3\mathcal{H}^{2}\psi + \frac{\delta{\Lambda} a^{2}}{2}= \nonumber \\
& \;\;\;\;  =  4\pi G_{0}a^{2}(\delta{T^{0}_{0}} +2 \nu\psi \pertorder{0}{T^{0}_{0}}) \, ,\qquad \label{EffEqs1}
\end{align}
\begin{align}
\partial_{i}\left[ {\phi}' + \nu{\psi}'+\mathcal{H}\psi(1- \nu) \right] = -4\pi G_{0} a^{2}\delta{T^{0}_{i}}\, , \label{EffEqs2}
\end{align}
\begin{align}
&\left[{\phi}''+ \nu {\psi}''+\mathcal{H}({\psi}'+2{\phi}'+ \nu{\psi}')+  \right. \label{EffEqs3} \\ 
& \;\;\;\; \left. +\frac 12 \nabla^{2}(\psi-\phi-2 \nu \psi)+\psi \left( 2\mathcal{H}'+\mathcal{H}^{2}\right)+ \frac{\delta{\Lambda} a^{2} }{2} \right]\delta^{i}_{j} - \nonumber \\
& \;\;\;\; -\frac 12 \partial_{j}\partial^i(\psi-\phi-2 \nu \psi) = 4\pi G_{0}a^{2}(\delta{ T^{i}_{j}}+ 
2 \nu\psi \pertorder{0}{T^{i}_{j}} )\, . \nonumber
\end{align}
In the above, a prime denotes derivative with respect to the conformal time $\eta$, we are no longer using ``$\stackrel*=$'' to emphasize the use of a particular reference frame, and all the computations are assumed to be up to the first perturbative order, even though there are no ``$O(\psi^2)$'' or ``$\approx$'' being used. All the computations will be exact on $\nu$, unless otherwise stated. This is important to uncover theoretical bounds for $\nu$ and for completeness. We also remark that the limit $\nu \rightarrow 0$ leads to GR.

From the non-diagonal part of eq.~\eqref{EffEqs3}, one infers the gravitational slip parameter as \cite{Amendola:2016saw,Pizzuti:2019wte, Ishak:2018his}
\begin{equation} \label{slip}
 \frac{\phi}{\psi} = 1-  2  \nu \, .
\end{equation}
Using this result, the field equations can be written in the Fourier space  as
\begin{align}
&3\mathcal{H}(1- \nu)\psi'+3\mathcal{H}^{2}\psi +8 \pi G_{0} \pertorder{0}{\epsilon} a^2\, \nu\,\psi+k^{2}(1-  \nu)\psi+ \nonumber \\ 
& \;\;\;\; +\frac{\delta{\Lambda} a^{2}}{2}  = -4\pi G_{0}a^{2}\delta\epsilon,\qquad \label{timetime}
\end{align}
\begin{eqnarray} \label{velocity}
(\psi'+\mathcal{H}\psi)(1- \nu) = 4\pi G_{0}a^{2}(\pertorder{0}{\varepsilon}+\pertorder{0}{p})\frac{\theta}{k^{2}} \,,
\end{eqnarray}
\begin{align} 
&(1- \nu)\psi''+3\mathcal{H}(1- \nu)\psi'+(2\mathcal{H}'+\mathcal{H}^{2})\psi- \nonumber \\
& \;\;\;\; - 8\pi G_{0} a^2 \pertorder{0}{p}  \nu\,\psi+ \frac{\delta{\Lambda} a^{2}}{2}= 4\pi G_{0}a^{2}\delta p \, . \label{spacespace}
\end{align}
In the above, eq.~\eqref{velocity} is the divergence of eq.~\eqref{EffEqs2} and it was introduced  $\theta \equiv \partial^{i}\vartheta_{i}$.  These equations also show that $ \nu = 1$ is a very special case, as it will be further detailed latter.

\subsection{Scalar perturbations solutions for particular cases} \label{sec:scalarpertsol}

The perturbative solution for a universe with $\Lambda_0 =0$ and with either dust or radiation can be presented in analytical form. From eqs.~(\ref{timetime}, \ref{spacespace}) with  $p = w \epsilon$ and constant $w$,  one finds
\begin{align} 
&\left [ \psi''+3\mathcal{H}(1+w)\psi'\right] (1- \nu)+ \nonumber \\
& \;\;\;\; +\left[w k^{2}(1- \nu)+ (1+ 3w)\mathcal{H}^{2}+ 2\mathcal{H}' + \right. \nonumber \\ 
& \;\;\;\; \left. + (1+w) \nu\Lambda_{0}a^2 -3 \nu(1+w)(\mathcal{H}'+\mathcal{H}^{2}) \right] \psi= 0\,. \label{master}
\end{align}

One can directly check that for $\omega = 1/3$ and negligible $\Lambda_0$ there will be no RG effects on $\psi$, that is, for radiation fluid the  solution for $\psi$ is the same of GR. The $\phi$ solution will also be equal to the GR solution, apart from a constant factor, which comes from the slip parameter. 

For the case of a universe with dust only, with  $w=0$ and $\Lambda_0=0$,  eq.~\eqref{master} becomes
\begin{eqnarray} \label{masterdust}
\left ( \psi''+3\mathcal{H}\psi' \right ) (1- \nu) - \frac 32  \nu \mathcal{H}^{2}\psi= 0\,,
\end{eqnarray}
where it was used that $\mathcal{H}' = -\frac 12 \mathcal{H}^2$. The solution reads
\begin{eqnarray}\label{soldust}
\psi = C_1 \, \eta^{\tilde \nu } + C_2\, \eta^{-\tilde \nu - 5} \,,
\end{eqnarray}
where $C_1$ and $C_2$ are integration constants with respect to $\eta$, they depend on the wavenumber $k$, and 
\begin{equation}
	\tilde \nu \equiv \frac 12  \left( \sqrt{1 + \frac{24}{1 - \nu}} - 5 \right ) \, .
\end{equation}
The constant $\tilde \nu$ is a shorthand notation and it is such that $\nu =0$ implies $\tilde \nu=0$. It is also a monotonous crescent function in the domain $-\infty < \nu < 1$, such that $-2 < \tilde \nu < \infty$. The $\psi$ term that depends on $C_2$ necessarily decays with time. The GR solution (i.e., $\psi = C_1  + C_2\eta^{-5}$) is reproduced in the limit $ \nu \rightarrow 0$.   The above solution is exact on $\nu$ and puts an upper bound on it, namely 
\begin{equation}
	 \nu < 1 \, .	 \label{nuExactbound}
\end{equation}
The case $\nu > 25$ also provides real results for $\psi$, but it has no GR limit and will not be further considered here.

Considering an expansion on $ \nu$ up to its first order, eq.~\eqref{soldust} becomes especially simple
\begin{eqnarray} \label{solphi}
\psi\approx C_1\left(1 +\frac{6}{5}  \nu \ln\,\eta \right)  + \frac{C_2}{\eta^5} \left( 1 - \frac 6 5 \nu \ln \eta \right)\,. \label{philog}
\end{eqnarray}
Hence, apart from the decaying mode (proportional to $C_2$), for a universe that is dust dominated, the first nontrivial RG correction is the introduction of a logarithm time dependence in the Newtonian potential, contrasting to the GR case of constant Newtonian potential for dust.

To summarize, for negligible $\Lambda_0$ and in a radiation dominated universe ($T = 0$) we find  no corrections on the $\psi$ solution with respect to GR. For a dust dominated universe and for $|\nu| \ll 1$, the single change in the $\psi$ solution  is the addition of a $\ln \eta$ term \eqref{philog}, thus providing a slow time variation of $\psi$ in a matter dominated universe.  For both cases, the $\phi$ solution is derived immediately from the slip \eqref{slip}, which is a constant that differs from the GR value of 1. This constant slip is already a clear difference with respect to both $f(R)$ and many scalar-tensor theories, including the Brans-Dicke case, since for the latter the slip parameter is not a constant.

\subsection{Density contrast evolution and the Jeans length} \label{sec:jeans}

This subsection aims to qualitatively explore the $G$ and $\Lambda$ running effects for structure formation.  The case of  interest here is that of a universe dominated by matter and with negligible influence of $\Lambda_0$. 

Besides the evolution of the perturbation $\psi$, it is also relevant to understand how matter perturbations evolve. From the time-time component of the perturbative field equations \eqref{timetime}, and using the $\psi$ solution for dust \eqref{soldust}, it is possible to find an explicit result for the density contrast $\delta_\varepsilon$, with
\begin{equation}
	\delta_\varepsilon \equiv \frac{\delta \varepsilon}{\varepsilon}\, .
\end{equation}
Using also the background solution for a dust universe with $\Lambda_0=0$ and the solution for $\Lambda$ \eqref{L1}, one finds
\begin{align}
	&\delta_\varepsilon  = \frac{C_1}{6} \eta^{\tilde \nu} \left[ -12\left (1 +\frac{\nu}{2} \right) - k^2 \eta^2(1- \nu) - 6 (1-\nu) \tilde \nu \right] + \nonumber \\
	& \! +\!  \frac{C_2}{6} \eta^{-\tilde \nu-5} \left [ 18 (1 - 2 \nu) - k^2 \eta^2(1 - \nu)  + 6 (1 - \nu) \tilde \nu\right ] . \label{deltaEpsilonExactOnNu}
\end{align}
The integration constants $C_1$ and $C_2$ above are the same that appear in the $\psi$ solution \eqref{soldust}. The GR solution is clearly recovered in the limit $\nu \rightarrow 0$ \cite{mukhanov2005physical}. If one assumes $\nu$ to be small, up to first order, one finds
\begin{align}
	&\delta_\epsilon \approx - \frac{C_1}{6} \left[  12 \left(1 + \nu \frac {11}{10}  + \nu \frac{6}{5}   \ln \eta \right)    \right .   \nonumber\\
	& \!+\! \left.  k^2 \eta^2 \left ( 1-\nu + \nu \frac {6}{5} \ln \eta \right ) \right]  \!+\! \frac{C_2}{6 \eta^5} \left [ 18 \left(1 - \frac{8}{5} \nu  - \frac {6}{5} \nu \ln \eta \right)\right .  \nonumber \\
	& \!+\! \left.  k^2 \eta^2 \left( -1 + \nu+ \frac 65 \nu \ln \eta \right) \right] \, . \label{deltaEsmallnu} 
\end{align} 
At subhorizon scales, and up to the first order on $\nu$, using eq.~\eqref{philog}, one can thus write
\begin{align}
	\delta_\varepsilon & \approx  \!-\! \frac{k^2 \eta^2}{6} \left[C_1 \left (1 - \nu + \nu \frac 65 \ln \eta \right) \!+\! \frac{C_2}{\eta^5}\left( 1-\nu - \frac 65 \nu \ln \eta \right) \right]  \nonumber \\ 
	& \approx - \frac 16  k^2 \eta^2 (1 - \nu) \psi     \, .
\end{align}
The relation between $\delta_\varepsilon$ and $\psi$ above is the same of GR, apart from the correction proportional to $\nu$. It could also be found from eq.~\eqref{deltaEpsilonExactOnNu} without the small $\nu$ approximation. At small scales, this $1-\nu$ factor is also found from a Jeans length analysis, as shown below. In the end, one can spot two corrections at small scales and up to the first order on $\nu$: $i$) the presence of the gravitational coupling correction $1-\nu$ and $ii$) the logarithmic   dependence on $\eta$ of $\psi$ and  $\delta_\varepsilon$.

To find the Jeans length, the first step we adopt is to find a second order differential equation for $\delta_\varepsilon$. Using $\nabla_\alpha T^{\alpha \beta}=Q^\beta$, $ p = w \epsilon$ (with constant $w$), $\delta p = c_s^2 \delta \epsilon$, and since  $Q^0$ is a first order quantity while $Q^i$ is zero, one finds, up to first order,\footnote{Apart from the $Q^\beta$ term, see for instance ref.~\cite{1107453984}.}

\begin{eqnarray} 
	&&\delta_\epsilon' + 3 \mathcal{H} (c^{2}_{s} - w) \delta_\epsilon = (1 + w) (3 \phi' - \theta) - \dfrac{ Q_0}{\pertorder{0}{\epsilon} }\, ,  \label{vel} \label{delta} \\[.1in]
  &&\theta' + \mathcal{H} (1 - 3w)  \theta = k^2 \left( \psi + \dfrac{c^{2}_{s}}{1 + w} \delta_\epsilon \right)\, . \label{thetaprime}
\end{eqnarray}

From eq.~\eqref{Qmatter} and using that $  \pertorder{0}{\varepsilon}' + 3 {\cal H} \pertorder{0}{\epsilon} =0$,
\begin{eqnarray} \label{deltaG}
 Q_0
&=&  - \, \delta_G  \, \pertorder{0}{\varepsilon}' = 6  \nu \psi \mathcal{H} \pertorder{0}{\epsilon}  \, .
\end{eqnarray}

By deriving eq.~\eqref{delta} and combining it with eqs.~(\ref{friedmann2}, \ref{slip}, \ref{thetaprime}, \ref{deltaG}),  it is possible to find a second order equation that governs the  density contrast dynamics. Such equation, in a matter dominated universe, takes the form
\begin{align} 
	& \delta_\epsilon'' +\mathcal{H} \delta_\epsilon' + \left (\frac 32 {\cal H}^2 + k^2 \right)c_s^2 \delta_\epsilon =  \label{pertm} \\
	& \;\;\;\; = 3(1-2  \nu)\psi'' + 3(1-4  \nu) \mathcal{H} \psi' - (k^2 + 3  \nu \mathcal{H}^2)\psi    \,.\nonumber 
\end{align}
In the above, we used $w=0$ and $c_s^2 \ll 1$.

 For computing the Jeans length, we are interested in the subhorizon limit and without neglecting $c_s$. Recalling that, in the subhorizon limit,  eq.~\eqref{velocity} implies $\psi' = - {\cal H} \psi$, then eq.~\eqref{pertm} can be written as
\begin{eqnarray} \label{deltas}
  \delta_\epsilon''+\mathcal{H}\delta_\epsilon'+\left(k^2c^2_s -\frac{4\pi G_0}{1- \nu } a^2  \pertorder{0}{\epsilon} \right)\delta_\epsilon =0 \,. \label{deltaEpsilonEvolution}
\end{eqnarray}

Therefore, the Jeans length is 
\begin{eqnarray} \label{jeans}
  \lambda_{\rm \small J} \equiv \frac{2 \pi a}{k_{\rm \small J} }= c_s \sqrt{\frac{(1- \nu ) \pi }{G_0 \pertorder{0}{\epsilon}}}  \,.
\end{eqnarray}
Hence, for small scales and for $0<  \nu < 1 $,  the RG effects  reduce $\lambda_{\rm \small J}$ and enhance the colapse of structures, while $ \nu <0$ decreases structure formation. Thus, the  ``force'' that acts on test particles is enhanced for $\nu>0$, similarly to \cite{Rodrigues:2009vf, Toniato:2017wmk}.

\subsection{Modified gravity parametrizations} \label{sec:modparametrizations}
Besides the slip parameter, another relevant parameter for describing cosmological models comes from the cosmologically extended Poisson equation (\ref{timetime}), and it is sometimes designated by $Q(a, k)$, where $Q$ is such that \cite{Clifton:2011jh, Amendola:2012ys, Amendola:2016saw}
\begin{eqnarray} 
-k^{2}\phi = 4\pi G_{0}Q(a,k)\,a^{2} \epsilon \, \Delta_\epsilon \, , \label{defQ}
\end{eqnarray}
with
\begin{equation}
	\Delta_\epsilon \equiv \delta_\epsilon + 3 (1 + w) {\cal H}\theta /k^2.
\end{equation}
If more than one fluid is being considered, then there should be a sum on $\delta_\epsilon$, $w$ and $\theta$.

From eqs.~\eqref{slip, timetime}, one finds
\begin{equation}
	Q(a,k) = \frac{1 - 2  \nu}{1 - \nu + \nu \, q (a)/k^2} \, ,
\end{equation}
with
\begin{equation}
	q(a) = 12 \pi G_0  (\pertorder{0}{\epsilon}+\pertorder{0}{p}) a^2	\, .
\end{equation}
The result above is independent on the value of $\Lambda_0$. We remark that the found expressions for the gravitational slip $\phi/\psi$ and $Q(a, k)$ are not common ones, in particular they differ from Brans-Dicke and $f(R)$ gravity expressions. 

We stress that $\pertorder{0}{p}$ and $\pertorder{0}{\epsilon}$ are background values for the pressure and energy density of the matter fields, they do not include $\Lambda_0$, hence $\pertorder{0}{p}\ll \pertorder{0}{\epsilon}$ does not impose any limit on $\Lambda_0$. Thus, for a universe with $\Lambda_0$, with matter and negligible radiation, 
\begin{align}
	q(a)|_{\pertorder{0}{p}\ll \pertorder{0}{\epsilon}} & = 12 \pi G_0 \pertorder{0}{\varepsilon}(\eta_0) \, a^{-1} = \frac{9}{2 a } \Omega_{m0} H_0^2 \nonumber \\
		& \approx \frac{1}{ a } \, ( 4 \, \mbox{Gpc})^{-2}\, . \label{qGpc}
\end{align}
Where $\pertorder{0}{\epsilon}(\eta_0)$ is the value of the background energy density at a time $\eta_0$ (today). The above estimate for $q(a)$ is based on the $\Lambda$CDM value for $\Omega_{m 0} H_0^2 $ with $a(\eta_0) =1$. Hence, as a function of the redshift $z$,
\begin{equation}
	Q(z,k) \approx \frac{ 1 - 2\nu}{1 - \nu + \nu  (1+z) / \left( 4  k {\rm Gpc} \right)^{2}} \,.
\end{equation}
For $z \approx 0$, the dependence on $k$ only becomes relevant for distances of the Gpc order or larger. For  $\nu >0 $, and $z \gg 1$, one sees that $Q$ decreases with $z$. This behaviour indicates  deviations from $\Lambda$CDM for the primordial universe, but only at the perturbative level. It should be  recalled that the above expression for $Q$ cannot be extended towards arbitrary $z$ values, since before the matter-radiation equality radiation pressure will not be negligible. And also, as  shown in section \ref{sec:scalarpertsol}, the $\psi$ solution for a radiation dominated universe is the same of GR. Although this behaviour of $Q(z,k)$ is interesting and should be further studied, here we will continue to focus on the universe at late times.   We also stress that the assumption of analyticity of $G(W)$ may  work as a good approximation within a given range for $W$, not necessarily for any $W$ value.

Since the derived gravitational slip is constant, it is trivial to convert the $Q$ result into an expression for $Y$, that is, the analogous quantity with $\phi$ replaced by $\psi$ in the left hand side in eq.~\eqref{defQ}. It reads
\begin{eqnarray}
	Y &=& \frac{1}{1 -   \nu +  \nu q(a)/k^2} = \frac{Q}{\phi/\psi} \, . \label{Yparameter}
\end{eqnarray}
And the lensing parameter, relevant for weak lensing and the integrated Sachs-Wolfe effect, reads \cite{Amendola:2012ys}
\begin{equation}
	\Sigma \!=\! \frac 12 Q \left(1 + \frac{\psi}{\phi} \right) \!=\! \frac{1 -   \nu}{1 -  \nu + \nu q(a)/k^2 }\,. \label{Sigmaparameter}
\end{equation}
Which shows that at distances smaller than one Gpc and for $z \lesssim 1$, $\Sigma$ does not depend on $\nu$ and it satisfies $\Sigma=1$, which is the same value of GR.

A set with two of the four parameters ($\phi/ \psi, Q, Y, \Sigma$) is sufficient for describing the dynamics of the first order perturbations for many modified gravity theories \cite{Amendola:2012ys}. However, we are considering here a framework in which the energy-momentum tensor is not always conserved, hence  the matter perturbations  may depend on two of the previous parameters and the $Q^\beta$ vector \eqref{Qvectordef}.

\subsection{Constraining $\nu$ from modified gravity parametrizations} \label{sec:constraintsnu}

The results of section \ref{sec:modparametrizations}  will be here used to constrain $\nu$. To this end, there are two issues to considered: $i)$ many constraints that can be found in the literature assume particular time and $k$ dependencies that do not match those here found;  $ii)$ energy-momentum conservation is commonly assumed in the literature, while in this framework it is in general violated.

Considering the item $ii$ above,  we note the following particularities of this specific case: $a$)~ energy-momentum tensor is always conserved at background level; $b$)~any energy-momentum with zero trace at background level is  conserved even at the first order; and $c$)~in a comoving frame with the cosmological background, particles follow geodesics (as shown in section \ref{sec:eomrel}). In particular this implies that, for a comoving observer and up to the first perturbative oder, the trajectory of light and that of massive isolated particles are the same of GR for a given metric.

The bounds on the gravitational slip proposed in refs.~\cite{Pizzuti:2016ouw, Pizzuti:2019wte} are based on a comparison between the potential $\psi$ inferred from the internal dynamics of clusters of galaxies with lensing effects from the same clusters.  The bounds from  ref.~\cite{Pizzuti:2016ouw} are not particularly strong, but are sufficient to yield $|1-\phi/\psi| \leq 0.61$ at 2$\sigma$ level (and apart from systematic errors), which implies, from \eqref{slip}, that  
\begin{equation}
	| \nu| \leq 0.30	 \label{boundnuclash}
\end{equation}
at $2\sigma$ level. A forecast considering near future surveys is done in \cite{Pizzuti:2019wte}, where it is found the stronger bound $|1-\phi/\psi| \leq 0.09$, at 2$\sigma$ level.  Consequently, 
\begin{equation} \label{boundnuPizzuti}
	| \nu| \leq 0.04 \, ,
\end{equation}
at 2$\sigma$ level. This is a significant constraint for the perturbations. The test above works in the following way: assuming that current observations are in agreement with $\Lambda$CDM, it states what could be the largest gravitational slip deviation from the fiducial value of 1 that would be still in agreement with observations.

One can find many other constraints in the literature with different hypothesis (e.g., \cite{Aghanim:2018eyx, Abbott:2018xao}), whose application to this framework requires the use of some approximations which may or may not be reasonable (for instance, on the redshift dependency). Nonetheless,  they imply constraints on $\nu$ for $z\approx 0$ about the same order of eq.~\eqref{boundnuclash}.

\subsection{On the cosmological dark matter and dark energy interpretation} \label{sec:dark}

Can the dynamical change provided by these RG corrections have a direct impact on dark matter at cosmological scales? Considering changes of the $10\%$ order on large  structures (say a dark matter filament or a large  cluster of galaxies) for the current bound \eqref{boundnuclash}, there is space for non-negligible ($\sim 10\%$) changes. These changes, however, would be present only as constant global enhancements of the dynamical effects, since the $Y$ parameter can be enhanced due to $\nu$ changes \eqref{Yparameter}. This is also closely related to the Jeans length rescaling  due to $\nu$, as shown in eq.~\eqref{jeans}. There would still be a large need for dark matter at cosmological scales even without considering the bound from eq.~\eqref{boundnuclash}, this since, apart the constant rescaling of $Y$, the only scale dependent effect happens at Gpc scales \eqref{qGpc}, while dark matter clumps at smaller scales.

While $\nu>0$ implies an enhancement of local gravitational attraction, at Gpc distances the $q(a)$ term in the $Y$ parameter may become relevant, and the effect of the latter is to decrease $Y$, thus reducing the gravitational attraction at large distances for positive $\nu$. Similarly to the dark matter case, it will not remove the need for $\Lambda_0$, but it may have a non-negligible impact. Since the background equations of this framework are that of $\Lambda$CDM, the  best-fit values of $\Lambda_0$ and $H_0$ will be the same considering background only observables. For the perturbations, due to the extra parameter $\nu$, larger error bars are expected, but without a complete numerical analysis, using CMB data, it is not yet possible to say if the best value for $H_0$ will be closer with respect to the background one.  This framework is in the end a variation with respect to $\Lambda$CDM, and it may have impact on some $\Lambda$CDM tensions \cite{Riess:2019cxk, DiValentino:2019dzu, Camarena:2019moy}. Further and more detailed tests, using in particular the CMB data, are necessary and constitute a work in progress.

\subsection{Consequences for $f \sigma_8$} \label{sec:fs8}

Since the background field equations of this framework are the same of $\Lambda$CDM, we have to look for observables sensitive to the perturbations. We did this with the modified gravity parametrizations evaluated in the previous subsections, which lead to bounds, but no explicit links towards  solving  some of the current $\Lambda$CDM anomalies. Although it is beyond the purpose of this work to do a complete cosmological analysis, we consider here  $f \sigma_8$ data, which have shown discrepancies  at low redshift ($z \lesssim 2$) (e.g., \cite{Macaulay:2013swa, Barros:2018efl, Skara:2019usd}) with the $\Lambda$CDM parameters as inferred from the {\it Planck} collaboration \cite{Aghanim:2018eyx}.

For a universe with dust and $\Lambda$, the density contrast second order evolution equation  can be written as 
\begin{equation}
	  \delta_\epsilon''+\mathcal{H}\delta_\epsilon' - 4\pi G_0 Y(a,k) a^2  \pertorder{0}{\epsilon} \delta_\epsilon =0 \, , \label{deltaEpsilonYEvolution}
\end{equation}
where $Y$ is given by eq.~\eqref{Yparameter}. For distance scales much smaller than 4 Gpc $a^{-1/2}$, the term $q(a)/k^2$ inside $Y$ is negligible, thus $Y \approx 1/(1-\nu)$. The resulting expression is exact on $\nu$ and it is equivalent to  eq.~\eqref{deltaEpsilonEvolution} for negligible $c_s^2$. Although there is no explicit dependence on $\Lambda_0$, its effect is present in the background quantities.

With respect to the scale factor $a$ and using the physical time Hubble parameter $H$, eq.~\eqref{deltaEpsilonYEvolution} can be written as \cite{Skara:2019usd} (see also \cite{EspositoFarese:2000ij, Nesseris:2017vor})
\begin{equation}
\partial^2_a\delta_\varepsilon+  \left(\frac{ \partial_a H}{ H}+\frac{3}{a}\right) \partial_a\delta_\varepsilon - \frac{3 H^2_0 \Omega_{\rm m0} }{ 2 a^5  {H}^2} \frac{1}{1-\nu }  \delta_\varepsilon =0 . \label{deltaaDiffEq}
\end{equation}
In the above, $\partial_a$ is a derivative with respect to the scale factor $a$, $H_0$ is the value of $H$ today and we have used that $Y \approx (1-\nu)^{-1}$. We stress here a particular feature of this framework at subhorizon scales: we note that it is not the change of $Y$ with $z$ that allows for a possible tension reduction between the CMB and ``local'' measurements of $\sigma_8$, but the mismatch between the background gravitational constant ($G_0$) and the effective gravitational constant ($Y G_0$) that act on the perturbations, this adds a new relevant parameter for the dynamics.

As usual, the Hubble parameter as a function of $a$ is
\begin{equation}
	H^2(a) =  \left( \frac{\Omega_{\rm m0 }}{a^{3}} + 1 - \Omega_{\rm m 0} \right ) H_0^2 \, .
\end{equation}
For this case, which considers the influence of $\Lambda_0$, we do not know of an explicit analytical solution for $\delta_\varepsilon$, but eq.~\eqref{deltaaDiffEq} can be used to provide a numerical solution.

The main quantity for this test is $f\sigma_8(a)$, which is given by 
\begin{align}
	&f(a)  = \frac{d \ln \delta_\varepsilon (a)}{d \ln a} \, ,\\[.1in]
	&\sigma(a) = \sigma_8 \frac{\delta_\varepsilon(a)}{\delta_\varepsilon (a=1)} \, . \\[.1in]
	& f\sigma_8 (a)  \equiv f(a) \sigma(a) =  \frac{\sigma_8}{\delta_\varepsilon (1)} a \partial_a \delta_\varepsilon(a) \, .
\end{align}

To solve eq.~\eqref{deltaaDiffEq} numerically, it is important to know the initial conditions.  Considering standard $\Lambda$CDM background, at $z\sim 10^3$ the universe is dominated by dust, hence using initial conditions at this $z$ from the dust-only solution \eqref{deltaEpsilonExactOnNu} should work as a good approximation in the $\Lambda$CDM context (i.e., $\nu=0$, see e.g., \cite{Skara:2019usd}). In order to better evaluate the impact of such approximation to the model under consideration ($\nu\not=0$), we also consider imposing initial conditions and using eq.~\eqref{deltaEpsilonExactOnNu} at $z=100$. We find that there are no relevant changes to any of our results using either one of the cases, the difference on the inferred parameters are about $\sim 10^{-4}$.

At subhorizon scales, only the terms that multiply $k^2 \eta^2$ are relevant. The mode proportional to $C_1$ will eventually dominate over the one proportional to $C_2$, hence we only consider the $C_1$ mode (which, apart from the case $\nu \ll -1$, it is an increasing mode, while $C_2$ is the coefficient of a decreasing mode). Neglecting the decreasing mode is commonly done (e.g., \cite{Skara:2019usd}) since it simplifies considerably the issue of initial conditions while the approximation is a very good one: indeed, at $z\sim 100$ one is deep in the matter dominated phase, thus providing sufficient time for the decreasing mode to be negligible.

For a dust dominated universe $a \propto \eta^2$, hence eq.~\eqref{deltaEpsilonExactOnNu} yields
\begin{equation}
 \delta_\varepsilon(a) \propto  a^{\frac{\tilde \nu}{2} + 1} \; \mbox{ and } \; \partial_a \delta_\varepsilon(a) \propto  \left(\frac{\tilde \nu}{2} + 1\right) a^{\frac{\tilde \nu}{2}}\, ,
\end{equation}
this in the subhorizon limit and with only the increasing mode. Since eq.~\eqref{deltaaDiffEq} is invariant under any constant rescaling of $\delta_\varepsilon$, only the $a$ dependent term is kept in $\delta_\varepsilon(a)$, while for deriving $\partial_a \delta_\varepsilon(a)$ we have not further rescaled $\delta_\varepsilon$. Within GR, for $z=10^3$, one would use the boundary conditions $\delta_\varepsilon(10^{-3}) = 10^{-3}$ and $\partial_a\delta_\varepsilon(10^{-3}) = 1$.

\begin{figure*}[hbt]
	\begin{center}
		\includegraphics[height=6.25cm]{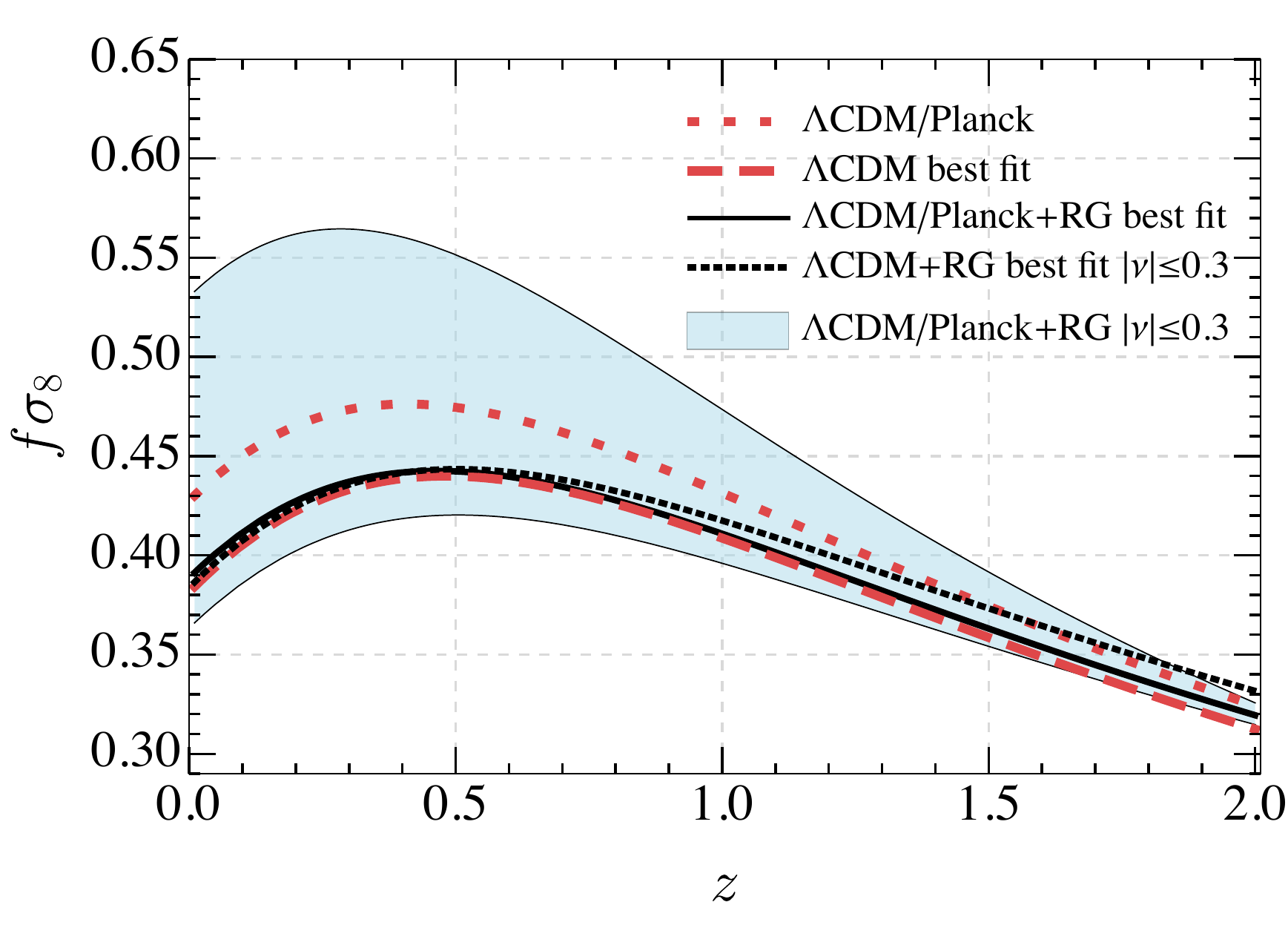}
		\includegraphics[height=6.25cm]{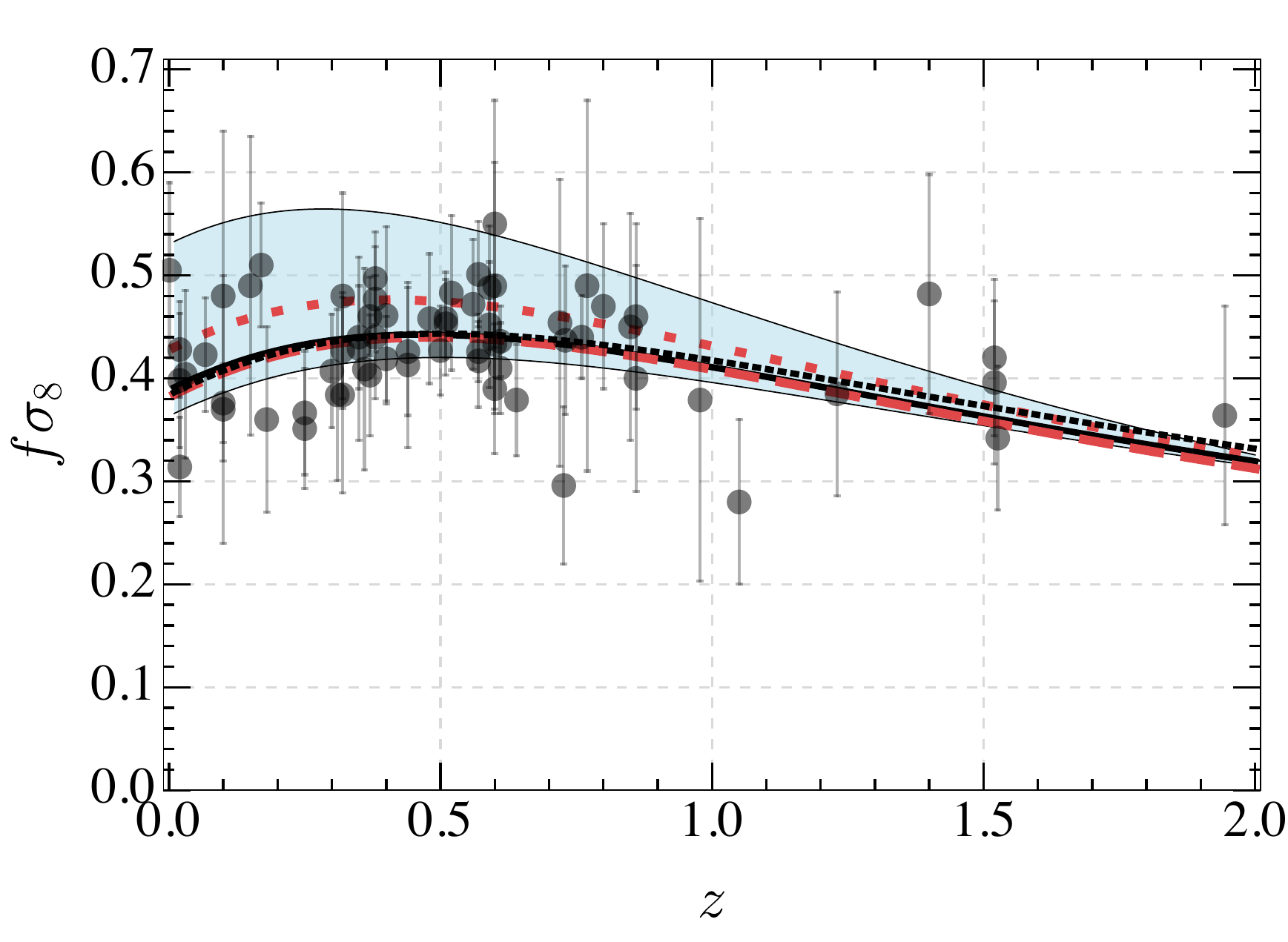}
		\caption{ {\bf $f\sigma_8$ model comparison.} {\it Left:} $f \sigma_8$ curves for four models that appear in Table \ref{tab:results}, only the  best fit model without constraints is not shown (since it violates the bound \eqref{boundnuclash}, and since its curve is similar to the dotted black curve). $\Lambda$CDM/Planck means that the parameters $\Omega_{\rm m0}$ and $\sigma_8$ are those given by the 2018 Planck collaboration \cite{Aghanim:2018eyx}. The bluish region shows the effect of changing the value of $\nu$, within the allowed bounds \eqref{boundnuclash}, while using the values of $\Omega_{m0}$ and $\sigma_8$ given by $\Lambda$CDM/Planck \cite{Aghanim:2018eyx}. {\it Right:} This plot shows the same curves displayed in the left plot and it adds the $f\sigma_8$ observational data (data compiled by ref.~\cite{Skara:2019usd}).}
		\label{fig:results}
	\end{center}
\end{figure*}

Following ref.~\cite{Skara:2019usd},  our results can be seen in Table \ref{tab:results} and in Fig. \ref{fig:results}.  The results  show that $\nu$ has a relevant impact on $f \sigma_8$ data, even within the bound \eqref{boundnuclash}, and therefore this framework may alleviate possible incompatibilities between $\sigma_8$ as inferred from the CMB with $\sigma_8$ values inferred at low redshift. In more detail, in Table~\ref{tab:results} we show the results for standard $\Lambda$CDM and consider its extended version with cosmological RG effects, as here proposed and labeled  as $\Lambda$CDM+RG. The simplest case here considered is that of  $\Lambda$CDM with parameters $\Omega_{\rm m0}$ and $\sigma_8$ fixed from the CMB \cite{Aghanim:2018eyx}, while $\nu$ is allowed to vary to better accommodate the model within the $f\sigma_8$ data (the third line in Table \ref{tab:results}). Clearly, $\nu$ has a relevant impact on this fit and the result is as good as (considering the value of $\chi^2_{\rm min}$) the case in which both $\Omega_{\rm m0}$ and $\sigma_8$ are allowed to vary within $\Lambda$CDM. The same table shows the case in which $\Omega_{\rm m 0}, \sigma_8 \mbox{ and } \nu$ are free to vary, which slightly further reduces the $\chi^2_{\rm min}$ value but the resulting $\nu$ value that is outside the bound \eqref{boundnuclash}. By constraining $\nu$ to lie within that bound, the resulting $\chi^2_{\rm min}$ changes by only 0.10 while the $\Omega_{\rm m0}$ and $\sigma_8$ become close to the $\Lambda$CDM/Planck values (as shown in the last line of Table~\ref{tab:results}). As previously commented, the impact on the amount of dark matter is a small one, it is far from replacing dark matter at cosmological scales; and actually these data suggest a slight increase on the dark matter content, as expected since these dada favour negative values of $\nu$.

In Fig.~\ref{fig:results}, we show the curves corresponding to four of the best-fit results presented in Table \ref{tab:results}. We only omit the case that violates the constraint \eqref{boundnuclash}. The plot also explicitly shows that a change of $\nu$  can have a sizable effect on $f\sigma_8$, and this especially for low $z$, even considering the constraint \eqref{boundnuclash}. For larger values of $z$ (i.e., $z \gtrsim 1.5$), different $\nu$ values lead essentially to the same predictions. This framework can be further tested by either extending this analysis towards full CMB data or by results from future gravitational-slip bounds. The latter can either further support this approach, or may render the possible effects of this framework on $f\sigma_8$ as a minor one, as implied by the forecasted bound \eqref{boundnuPizzuti}.

In ref.~\cite{Sola:2019jek}, the authors find particular Brans-Dicke gravity solutions that can mimic certain RG corrections to gravity \cite{Peracaula:2018dkg,Perez:2018qgw} and alleviate both the $H_0$ and $f\sigma_8$ tensions.\footnote{The Running Vacuum model, considered in refs.~\cite{Peracaula:2018dkg,Perez:2018qgw}, is related to a class of RG corrections to gravity \cite{Sola:2013fka}.} In their case, the background field equations are different from $\Lambda$CDM, which is different from our case, but, on the other hand, at subhorizon the main new effect is a rescaling of the effective gravitational constant. Both models introduce a departure from GR that favour cosmological perturbations whose effective gravitational constant is reduced with respect to that of GR.

\begin{table*}
\caption{Best-fit results for $f \sigma_8$ data.}
\centering
\renewcommand{\arraystretch}{1.5}
\begin{tabular}{llllrrrr}
\hline\hline
Model & Variation & fitted parameters & $\chi^2_{\rm min}$ & $\Omega_{\rm m0}$ & $\sigma_8$ & $\nu$ \\
\hline
$\Lambda$CDM & Planck-2018 parameters & None & 51.34 & 0.315 & 0.811 & 0\\
$\Lambda$CDM & best fit from $f\sigma_8$ data & $\Omega_{\rm m0}$, $\sigma_8$ & 32.40 & 0.283 & 0.769  & 0\\
$\Lambda$CDM+RG & Only $\nu$ is fitted   & $\nu$& 32.42 & 0.315  & 0.811 & $-0.167$ \\
$\Lambda$CDM+RG & best fit, no constraints & $\Omega_{\rm m0}$, $\sigma_8$, $\nu$ & 32.04 & 0.355 & 0.981  & $-0.769$ \\
$\Lambda$CDM+RG & best-fit with $|\nu| \leq 0.3$  & $\Omega_{\rm m0}$, $\sigma_8$, $\nu$& 32.14 & 0.316  & 0.855  & $-0.300$ \\
\hline \hline
\end{tabular}
\label{tab:results}
\end{table*}

\section{Conclusions} \label{sec:conclusions}

Here we  presented  cosmological implications from scale-dependent couplings $G$ and $\Lambda$, considering that all the information on their running is included in the action. This approach is motivated from possible renormalization group (RG) effects to general relativity (GR) at large distances, together with the assumption that there should be an effective classical action capable of fully describing the dynamics at cosmological distances.  After presenting the full action, which extends that of ref.~\cite{Rodrigues:2015hba} by using an arbitrary number of possible RG scales, the field equations and consequences for the energy-momentum conservation are evaluated.  In the context of fluids, we use the same  RG scale proposed in ref.~\cite{Rodrigues:2015hba}, which is the $W$ scalar \eqref{Wdef}. The latter scale extends a number of noncovariant proposals (e.g., \cite{Reuter:2003ca, Shapiro:2004ch, Reuter:2004nx, Rodrigues:2009vf}).  This scale always preserves the background and affects the perturbations. The novelty in the cosmological case is that the background itself is dynamical, and hence, as here found, this property implies that a second RG scale is necessary. However, there is no second choice to be done, the field equations  fix the second scale as a scalar function of the energy-momentum tensor trace. No other scales beyond the second one are necessary. For vacuum ($T_{\alpha \beta} \rightarrow 0$), the relation between $G$ and $\Lambda$ is the same of ref.~\cite{Rodrigues:2015hba} (i.e., $\Lambda \propto G^{-1}$) (see also \cite{Bonanno:2001hi, Bentivegna:2003rr, Bonanno:2004ki}). After several dynamical consequences are detailed, including some exact solutions and  the coupling to fluids at the action level, the evolution of the first order perturbations are parametrized using $\phi/\psi, \Sigma, Y$ and $Q$, which are commonly used parametrizations to describe modified gravity (e.g., \cite{Amendola:2016saw}). From them, a clearer understanding of the cosmological effects from this framework is found, and bounds on the dimensionless $\nu$ parameter could be established. Our results are not compatible with the removal of either dark matter or dark energy in place of RG effects, but this framework can have relevant impact to both of them and possibly to anomalies at cosmological level \cite{Riess:2019cxk, DiValentino:2019dzu, Camarena:2019moy}. Numerical analysis  on the CMB power spectrum constitute a relevant piece of information for  addressing this issue, which is a work in progress. Further developments on the theoretical side, as a Hamiltonian formulation (e.g., \cite{Rodrigues:2018ioe}), are also being considered.

Our analysis on $f\sigma_8$ (sec.~\ref{sec:fs8}) shows that this framework can improve $\Lambda$CDM in this context if negative $\nu$ values are considered. This can appreciably reduce the  $f\sigma_8$ values for low redshift ($z<1.5$), while essentially preserving the $\Lambda$CDM $f\sigma_8$ results for higher redshifts (Fig.~\ref{fig:results}). Hence, it has the potential of alleviating tensions that are present in $\Lambda$CDM (e.g., \cite{Macaulay:2013swa, Barros:2018efl, Skara:2019usd}). Further analyses that consider more observational data  together are  still necessary.

The results here presented are not restricted by too specific assumptions on a $\beta$-function realization. The assumption is that   $G(\mu_1(W)) = G(W)$ can be approximated by a linear function about $W=0$ (higher order corrections are possible,  but do not change our results). One particularly relevant case, which includes a logarithm term, was discussed in detail in sec.~\ref{sec:eqsslip}. 

There are  different frameworks on scale-dependent couplings $\Lambda$ and $G$ at cosmological level. The majority considers the implementation of RG-like effects at the level of the field equations, or partially at the action level,   (e.g., \cite{Bertolami:1993mh, Reuter:2003ca, Grande:2010vg, Sola:2013fka, Hernandez-Arboleda:2018qdo, AgudeloRuiz:2019nnm}).\footnote{There are examples of models whose dynamics are similar to RG corrections at the field equations level, and that at cosmological scales the corrections can be written as powers on $H$, but they can be motivated from string theory (e.g., see \cite{Basilakos:2020qmu} and references therein).} Here we presented a framework in which all the relevant information come from the action, including the scale settings, and applied it to cosmology; leading to a picture that is different from both GR and well known modified gravity theories, as $f(R)$. Contrary to several approaches within RG effects at the field equations level, in the proposed framework, one cannot choose if either $G$ will vary, or $\Lambda$ will, or both of them (e.g., \cite{Grande:2010vg, Sola:2013fka}). This freedom appear in such theories since there the complete action is neither used nor known, hence $\nabla_\alpha T^{\alpha \beta}$ is not fixed and may either be zero or it may depend on the running of $G$ or $\Lambda$. In the framework here proposed (like that in ref.~\cite{Rodrigues:2015hba}) there is no such freedom, once the RG scales are fixed, $\nabla_\alpha T^{\alpha \beta}$ is also fixed. This is in accordance with eq.~\eqref{diff}, which is a consequence of diffeomorphism invariance of the action. 

\begin{acknowledgements}
We thank Ilya Shapiro for discussions and for commenting on a previous version of this work. We also thank Benjamin Koch for commenting on a previous version of this work. We thank the referee for relevant suggestions that improved this work. NRB thanks CAPES (Brazil) for support. FMS thanks FAPES (Brazil) for support. WSHR and DCR thanks CNPq (Brazil) and FAPES (Brazil) for partial financial support.   ``This study was financed in part by the {\it Coordena\c{c}\~ao de Aperfei\c{c}oamento de Pessoal de N\'ivel Superior - Brasil} (CAPES) - Finance Code 001.''	
\end{acknowledgements}

\end{document}